\newcommand{\dis}{\displaystyle} 
\newcommand{\mc}{\cal}
\documentstyle[12pt]{article}

\begin{document}
\parindent 1.3cm
\thispagestyle{empty}   
\vspace*{-4cm}
\noindent

\def\arccot{\mathop{\rm arccot}\nolimits}
\def\sd{\strut\displaystyle}

\begin{obeylines}
\begin{flushright}
UG-FT-101/99
hep-ph/9908299
August 1999
\end{flushright}
\end{obeylines}

\vspace{0.8cm}

\begin{center}
\begin{bf}
\noindent
FOUR FERMION CONTACT TERMS IN CHARGED CURRENT PROCESSES AND LARGE EXTRA DIMENSIONS
\end{bf}
  \vspace{1.2cm}\\
FERNANDO CORNET  \footnote{E-mail address: cornet@ugr.es}
\vspace{0.1cm}\\
Departamento de F\'\i sica Te\'orica y del Cosmos,\\
Universidad de Granada, 18071 Granada, Spain \vspace*{0.2cm}\\
MONICA RELA\~NO \footnote{E-mail address: mpastor@ll.iac.es} \vspace{0.1cm} \\
Instituto de Astrof{\'\i}sica de Canarias \\
E-38200 La Laguna, Tenerife, Spain \vspace*{0.2cm} \\
and \vspace*{0.2cm} \\
JAVIER RICO \footnote{E-mail address: javier.rico@cern.ch }
\vspace{0.1cm} \\
Institut f\"ur Teilchenphysik\\
ETHZ, CH-8093 Z\"urich, Switzerland
   \vspace{2.0cm}

{\bf ABSTRACT}

\parbox[t]{12cm}{ We study the bounds that can be obtained on four-fermion contact
terms from the experimental data for $e^+ p \to \overline{\nu} X$ obtained at HERA
and $p \overline{p} \to e^\pm \; \nu^{^{\hbox{\hspace*{-3mm}{\tiny (---)}}}}$,
measured at TEVATRON. We compare these bounds with the ones available in the
literature. Finally, we apply these results to study the compactification
radius in theories with large extra dimensions and we obtain the bound
$M_c \ge 3.3 \; TeV$ 
.}
\end{center}

\newpage

The Standard Model is in excellent shape from the experimental point of view. The
only problem appears to be the recently reported evidence for neutrino oscillations
\cite{OSCILATIONS}, but the agreement between the theoretical predictions and all the
high energy experimental data is remarkably good \cite{VANCOUVER}. Two years ago
an excess of events in neutral ($e^+ p \to e^+ X$) and charged current
($e^+ p \to \nu_e X$) Deep Inelastic Scattering was reported by the two HERA
experiments: H1 \cite{H1.1} and ZEUS \cite{ZEUS.1}. However, after an spectacular
increase in the collected luminosity this possible hint for Physics Beyond the
Standard Model has disappeared \cite{VANCOUVER2, MORIOND}. It is now interesting
(although certainly less exciting) to study the bounds that can be obtained from the
new data on the mechanisms that were proposed as possible explanations for the excess 
of events.

The effects of new physics can be parametrized in terms of higher dimension
($d > 4$) operators \cite{BUCHMULLER}. In particular, the first operators 
contributing to Deep Inelastic Scattering are dimension $6$, four-fermion 
contact interactions. These terms were introduced as an effective interaction
relevant in case quarks and leptons were composite objects \cite{PESKIN,RUCKL}.
However, it is clear that contact terms also appear as the low energy limit
of the exchange of heavy particles, in the same way as the $W^\pm$ gauge
boson exchange can be parametrized in the Fermi Lagrangian for energies much
lower than $M_W$. The main difference between both approaches to contact terms 
is the interpretation of the mass scale $\Lambda$ they contain. In the first case, 
compositeness, the mass scale is related to the inverse of the size of the 
composite object. In the second case, $\Lambda$ is related to the mass and coupling
constants of the exchanged particle. 

A particularly interesting application of contact terms appears as a result
of recent advances in superstring 
theories, where it has been observed that compact dimensions of a radius 
of $O(1 \; TeV^{-1})$ can be at the origin of supersymmetry breaking \cite{ANTONIADIS}.
Also, even larger compact dimensions, with sub-millimeter compactification
radius, allow to reduce the Planck scale to become of the order of a few $TeV$,
avoiding the gauge hierarchy problem \cite{ARKANI-HAMED}. So, if we have
$6-n$ dimensions of $O(TeV^{-1})$ and $n \ge 2$ dimensions with a compactification
size $O(mm - fm)$, in Type I/I$^\prime$ and Type IIA superstring theories we can
achieve that the scale at which gravity becomes strong and the string scale are,
both, $O(TeV)$ \cite{ANTONIADIS2}. Gravitons propagate in the $n$-dimensional space,
giving rise to new, effective interactions among the ordinary particles that live
in our conventional four dimensional space-time \cite{GIUDICE}. Gauge bosons 
propagate in the
$10 - n$ dimensional space and in the ordinary four dimensional space appear as
a Kaluza-Klein tower of states. Neutral current processes receive two types of new
contributions: graviton exchange and KK-states exchange, while charged current
processes only receive the second type of contributions. The phenomenology of these 
models is being discussed extensively in these days \cite{MUCHOS}. 

In this note we are going to use  recent data from HERA 
($e^+ p \to \overline{\nu}_e X$) 
and TEVATRON ($p \overline{p} \to e \overline{\nu}_e X$) to obtain bounds on the 
mass scale appearing in the $e \nu q q^\prime$ contact term and we will compare them
with the bounds obtained from the unitarity of the CKM matrix \cite{HAGIWARA}.
We will finish with a discussion of the connection of our results with extra
dimensions physics.

Low energy effects of physics beyond the SM, characterized by a
mass scale $\Lambda$ much larger than the Fermi scale, can be studied by a
non-renormalizable effective lagrangian, in which all the operators are
organized according to their dimensionality. Since the energies and momenta
that can be reached in present experiments are much lower than $\Lambda$, it
is expected that the lowest dimension operators provide the dominant
corrections to the SM predictions. Requiring $SU(2) \times U(1)$ invariance,
the relevant lagrangian for $e \nu q q^\prime$
charged current processes including dimension $6$ four-fermion operators is:
\begin{equation}
\label{LAG}
{\mc L} = {\mc L}_{SM} +  
     \eta^{lq} (\bar{l}\gamma_{\mu}\tau^{I}l) (\bar{q}\gamma^{\mu}\tau^{I}q)  + 
     {\mc L}_S,
\end{equation}
where ${\mc L}_{SM}$ is the SM lagrangian, $l=(\nu,e)$ and $q=(u,d)$ are the 
$SU(2)$ doublets containing the left-handed lepton and quark fields,
$\tau^I$ are the Pauli matrices and
${\mc L}_S$ are four fermion terms containing scalar currents instead of the
vector currents shown explicitly in Eq. (\ref{LAG}).  
It is customary to replace the coefficient $\eta$ by a mass scale $\Lambda$:
\begin{equation}
\eta = {\epsilon g^2 \over \Lambda^2},
\end{equation}
with $\epsilon = \pm 1$ taking into account the two possible interference 
patterns. For historical reasons $\Lambda$ is usually interpreted as the mass 
scale for new physics in the strong coupling regime, i.e. with
\begin{equation}
{g^2 \over 4 \pi} = 1.
\end{equation}
If the contact term is due to the $s$ or $t$ channel exchange of a heavy particle 
with a mass $M_h$ much larger than the center of mass energy then
\begin{equation}
\eta =  {\overline{g}^2 \over M_h^2} \; \; \hbox{and}  \; \; 
\Lambda = {\sqrt{4 \pi} M_h \over \overline{g}},
\end{equation}
where ${\overline{g}}$ is the coupling constant of the heavy particle to
a fermion pair.

We will not consider the terms in ${\cal L}_S$ in our analysis because it has been shown
that the ratio 
\begin{equation}
R = {\Gamma(\pi^\pm \to e^\pm \nu) \over \Gamma(\pi^\pm \to \mu^\pm \nu)} = 
(1.230 \pm 0.004) \times 10^{-4}
\end{equation}
provides a very strong bound on these terms:
\begin{equation}
\Lambda_s > 500 \; TeV.
\end{equation}
This is due to the fact that the scalar
currents appearing in  ${\cal L}_S$ do not lead to helicity suppression in the pion
decay amplitude as $V-A$ currents do \cite{ALTARELLI1,SHANKER}. 

The expression for the cross-section  $\sigma(e^+ p \to \overline{\nu} X)$ 
can be found in Ref. \cite{NOSOTROS}, where a first analysis 
to the HERA data was presented. Here we perform a fit to the combined data for
$d\sigma/dx dQ^2$ shown by the two
experiments, H1 and ZEUS \cite{VANCOUVER2,MORIOND}, at the Moriond and 
Vancouver conferences. The total integrated luminosity collected is 
$37 \; pb^{-1}$ for H1 and $47 \; pb^{-1}$ for ZEUS.
We have used the MRSA parameterization \cite{MRSA}, but
one cannot expect important changes in the result when using another 
parameterization. In our fit we have included Standard Model radiative corrections,
but we have neglected the interference between these corrections and the new
terms. Certainly, the data are compatible with the Standard Model
predictions and we can only obtain $95 \% C.L.$ bounds on $\Lambda^{lq}_+$ and
$\Lambda^{lq}_-$, where the subscript refers to the value of $\epsilon$. 
In order to obtain these bounds we have assumed that the 
probability density has the form:
\begin{equation}
f(z,n) = {z^{n/2 - 1} e^{-z/2} \over 2^{n/2} \Gamma(n/2)},
\end{equation}
where $n = m-1$, and $m$ is the number of data points included in the
fit. This expression corresponds to a $\chi^2$ distribution with $n$
degrees of freedom. The $95 \% C.L.$ bounds we have obtained are:
\begin{equation}
\label{HERA3F}
\begin{array}{lll}
\Lambda^{lq}_+ \geq 3.5 \; TeV & \hbox{and} & \Lambda^{lq}_- \geq 3.1 \; TeV.
\end{array}
\end{equation} 
It is interesting to note at this point that, due to the dominance of the
first family quarks in the proton structure functions, the results shown in
Eq. (\ref{HERA3F}) are strongly dominated by contact terms involving only
first family quarks and leptons, i.e. an $e \nu u d$ contact term. Indeed,    
neglecting new terms involving quarks from the second family we arrive to very
similar bounds:
\begin{equation}
\label{HERA1F}
\begin{array}{lll}
\Lambda^{lq}_+ \geq 3.2 \; TeV & \hbox{and} & \Lambda^{lq}_- \geq 2.8 \; TeV.
\end{array}
\end{equation} 

We now turn our attention to the closely related processes
$p \overline{p} \to e^\pm \; \nu^{^{\hbox{\hspace*{-3mm}{\tiny (---)}}}}$
measured at TEVATRON. At the partonic level this process is related to
the one at HERA via $t$ to $s$ channel exchange. This, however, introduces
a problem because the $W^\pm$ gauge boson can now be produced on-mass shell,
producing a very large background to study new physics. In our fit we have used the
data for $\dis{d\sigma \over dm_t}$, where $m_t$ is the $e \nu$ transverse mass,
from Ref. \cite{GERBER} with $m_t \geq 110 \; GeV$. This value has been chosen to 
optimize the sensitivity to new physics. The values for the $W$ mass 
and width we have used are:
$M_W = (80.41 \pm 0.10)  \; GeV$ and $\Gamma_W = (2.06 \pm 0.06) \; GeV$ \cite{PDG}
and we have checked that our results are not sensitive to changes
of these parameters within one standard deviation. The bounds we obtain,
\begin{equation}
\begin{array}{lll}
\Lambda^{lq}_+ \geq 2.0 \; TeV & \hbox{and} & \Lambda^{lq}_- \geq 1.2 \; TeV,
\end{array}
\end{equation} 
are much less restrictive than the ones obtained at HERA.

The most stringent bounds on lepton-quark charged current contact terms have
been obtained from the observed unitarity of the Cabbibbo-Kobayashi-Maskawa
matrix elements in Ref. \cite{KAORU}:
\begin{equation}
\label{UNITARITY}
|V_{ud}|^2 + |V_{us}|^2 + |V_{ub}|^2 = 0.9965 \pm 0.0021.
\end{equation}
Since CKM matrix elements are experimentally determined from the ratio of 
semileptonic to leptonic processes, both, lepton-quark and purely leptonic
contact terms contribute to Eq. (\ref{UNITARITY}):
\begin{equation}
\label{CKM}
V_{ud_j}^{obs} = V_{ud_j}^{SM}  
\left(1 - \dis{\eta^{lq} - \eta^{ll} \over 8 \sqrt{2} G_F} \right),
\end{equation}
where $\eta^{lq}$ and $\eta^{ll}$ stand for the lepton-quark and purely leptonic,
respectively, contact term couplings. The bounds on $\eta^{lq}$ and, consequently, on
$\Lambda^{lq}$ depend on the ones that can be obtained for $\eta^{ll}$ (or
$\Lambda^{ll}$). Hagiwara and Matsumoto have performed a fit to electroweak
parameters measured at LEP1, TEVATRON and LEP2 to obtain a value for
the $T$ parameter from which the bounds \cite{KAORU}:
\begin{equation} 
\label{LEPTONIC} 
\begin{array}{lll}
\Lambda_+^{ll} \geq 7.5 \; TeV & \hbox{and} & \Lambda_-^{ll} \geq 10.2 \; TeV
\end{array}
\end{equation}
were obtained. Introducing this result into Eqs. (\ref{UNITARITY}), (\ref{CKM})
and assuming that the contact terms are the same for all three families they 
obtained:
\begin{equation}
\label{BCKM}
\begin{array}{lll}
\Lambda_+^{lq} \geq 5.8 \; TeV & \hbox{and} & \Lambda_-^{lq} \geq 10.1 \; TeV.
\end{array}
\end{equation}
These bounds are more stringent than the ones obtained from HERA. However, one should
notice that both sets of bounds are complementary because of the different
assumptions used. Indeed,
the result (\ref{BCKM}) rely on the assumption that $V_{ud}$, $V_{us}$ and
$V_{ub}$ receive the same contribution from contact terms,
while the bounds obtained from ZEUS and H1 data are independent from this assumption, 
as we have explicitly shown in Eqs. (\ref{HERA3F}) and (\ref{HERA1F}). We should also
point out that the HERA data we used in our fit has been obtained with positron beams.
Since positrons interact via charged current processes with $d$ and $\overline{u}$
quarks in the proton, while electrons interact with $u$ and $\overline{d}$ quarks,
the cross section with positrons in the initial state is much smaller at large $x$ and
$Q^2$ than the one with electrons. Thus, with the same integrated luminosity
using electron beams as the one collected up to now with positron beams
the bounds shown in Eqs. (\ref{HERA3F}) and (\ref{HERA1F}) will
improve in such a way that for $\Lambda^{lq}_+$ can become similar to the one in 
Eq. (\ref{BCKM}).

The contact terms we have been studying up to now can be easily related to the
exchange of a tower of KK states corresponding to the $W$ boson. Such a tower
appears when the number of space-time dimensions is larger than $4$ and gauge
bosons can propagate in the new dimensions. For energies lower than the inverse
of the compactification radius $(R \sim 1/M_c)$ the gauge bosons propagating 
in the new dimensions appear as a tower of states with the same couplings as
the standard bosons. The lightest of the new states has a mass $O(M_c)$. For
$M_c \ge O(1 \; TeV)$ it is, thus, justified to approximate the effects of the
exchange of the new particles in four fermion processes by a contact term of the 
type introduced in Eq. (\ref{LAG}). 

The relation between the mass scale $\Lambda$ and the compactification scale
$M_c$ is particularly simple in the case of only one extra dimension with
compactification scale $O(TeV)$:
\begin{equation}
M_c^2 = { g^2 \Lambda^2_- \over 2 \pi} \sum_{n=1}^{\infty} {1 \over n^2},
\end{equation}
where $g$ is the $SU(2)$ coupling constant and the sum covers the
contribution of the infinite number of states. This sum s finite and
turns out to be $\pi^2/6$. In case there are more than one extra 
dimension with the same compactification radius, the sum is
divergent and a new parameter, a cut-off, must be introduced.
Since the coupling of the new states to
left handed fermions is universal, as it is the case for the $W$ boson, we just have 
take the most stringent bound for $\Lambda_-$ and convert it into a bound for $M_c$.
Thus, using $\Lambda_- \ge 10.2 \; TeV$ we obtain
\begin{equation}
M_c \ge 3.3 \; TeV.
\end{equation}

In summary, we have obtained bounds on the mass scale of the $SU(2) \times U(1)$ 
invariant, four-fermion, charged current contact term from the recent 
HERA and TEVATRON data. The first ones appear to be more sensitive to the 
presence of this contact term, but more luminosity (especially with electron 
beams) is needed before the bounds obtained from these processes can be competitive
with the ones obtained from the unitarity of the CKM matrix. Finally, we
have converted these results  into a bound on the compactification scale
of a large extra dimension in which the $W$ boson can propagate, obtaining
$M_c \ge 3.3 \; TeV$. This result is particularly interesting, not only because it is
one of the largest lower bounds obtained, but also because,being obtained form charged 
current processes, it is free from any
assumption on the effects of the KK tower of the graviton in theories with
a low gravity scale.

We thank  A. Kotwal, K. Hagiwara, W. Hollik and M. Masip for very helpful
discussions and comments. This research was partially supported by
CICYT, under contract number AEN96-1672, and Junta de Andalucia, under contract
FQM 101.

\end{document}